\documentclass[aps,pre,preprint,groupedaddress,showpacs]{revtex4}
\usepackage{amssymb}

\usepackage[dvips]{graphicx}
\usepackage{textcomp}
\usepackage{amssymb}

\begin{document}

\title{
\Large\bf Universal amplitude ratios in \\
the three-dimensional Ising universality class}

\author{Martin Hasenbusch}
\email[]{Martin.Hasenbusch@physik.hu-berlin.de}
\affiliation{
Institut f\"ur Physik, Humboldt-Universit\"at zu Berlin,
Newtonstr. 15, 12489 Berlin, Germany}

\date{\today}

\begin{abstract}
We compute a number of universal amplitude ratios in the 
three-dimensional Ising universality class. To this end, 
we perform Monte Carlo simulations of the improved Blume-Capel model 
on the simple cubic lattice. For example, we obtain $A_+/A_-=0.536(2)$
and $C_+/C_-=4.713(7)$, where $A_{\pm}$ and $C_{\pm}$ are the amplitudes
of the specific heat and the magnetic susceptibility, respectively. 
The subscripts $+$ and $-$ indicate the high and the low temperature phase, 
respectively. We compare our results with those  
obtained from previous Monte Carlo simulations, high and low 
temperature series expansions, field theoretic methods and 
experiments.
\end{abstract}
\pacs{05.50.+q, 05.70.Jk, 64.60.F-}
\keywords{}
\maketitle

\section{Introduction}
In the neighborhood of a second order phase transition various quantities 
diverge, following power laws. For example, in a magnetic system, 
the correlation length $\xi$, the magnetic susceptibility $\chi$
and the specific heat $C$  behave as 
\begin{equation}
\label{power}
 \xi \simeq f_{\pm} |t|^{-\nu}   \;\;\;\;,  \;\;
 \chi  \simeq C_{\pm} |t|^{-\gamma} \;\;\;\;,  \;\;  C \simeq A_{\pm} |t|^{-\alpha}
\;\;, 
\end{equation}
where $t=(T-T_c)/T_c$ is the reduced temperature.
The symbol $\simeq$ means asymptotically equal; corrections vanish
as $t \rightarrow 0$.
Critical exponents like $\nu$, $\gamma$ and $\alpha$ are universal.
That means, they take exactly the same value for all systems in a 
given universality class. A universality class is characterized
by the spacial dimension of the system, the range of the interaction
and the symmetry of the order parameter. For reviews on  critical phenomena
and its modern theory, the renormalization group see for example
\cite{WiKo,Fisher74,Fisher98,PeVi02}.

While individual amplitudes like $f_+$, $f_-$, $C_+$, $C_-$, $A_+$ and $A_-$
depend on the details of the system, amplitude ratios like 
$f_+/f_-$, $C_+/C_-$ and $A_+/A_-$ are universal.
The indices $+$ and $-$ indicate the high and the low
temperature phase, respectively. In addition to these simple ratios, there
are also more complicated universal combinations of amplitudes. The
combinations of the corresponding quantities are dimensionless. This means
that they have a combined critical exponent that is equal to zero.
Such amplitude ratios have been determined for a number of experimental systems
and computed by using various theoretical approaches like the 
$\epsilon$-expansion,
perturbation theory in three dimensions fixed, high and low temperature series 
expansions and Monte Carlo simulations. A summary of results is given in refs. 
\cite{ahp,PeVi02}.
Here we study universal amplitude ratios in the universality class of
the three-dimensional Ising model with short range interactions, which 
is characterized by the $\mathbb{Z}_2$ symmetry of the order parameter. 
This universality class is supposed to be realized in a huge range of
experimental systems: binary mixtures, uniaxial magnets or micellar
systems; see \cite{ahp,PeVi02}.

At finite values of the reduced temperature, power laws~(\ref{power}) are 
subject to corrections. For example the magnetic susceptibility behaves as
\begin{equation}
\label{chicorrection}
 \chi = C_{\pm}  |t|^{-\gamma} \left(1 + a_{\pm} |t|^{\theta} + b t + ...\right)
\;\;, 
\end{equation}
where $\theta = \nu \omega = 0.524(4)$ \cite{mycritical}. The amplitudes
$C_{\pm}$, $a_{\pm}$ and $b$ in general depend on the parameters of the system.
Already in 1982 the authors of ref. \cite{ChFiNi} have demonstrated that for
a model that interpolates between the Gaussian and the Ising model 
there is one value of the interpolation parameter, where $a_{\pm}$ vanishes. 
Renormalization group predicts that the zero of leading correction
amplitudes is the same for all quantities. 
In the following we shall call a model with $a_{\pm}=0$ an improved model.
Studying improved models simplifies the accurate determination of 
amplitude ratios using Monte Carlo simulations or high and low temperature
series expansions.
Here we simulate the improved Blume-Capel model on the simple cubic 
lattice. For the definition of this model see the next section.
Our main motivation to perform these simulations was to compute the energy
density of the bulk system in a large range of inverse temperatures. This
quantity is needed in our ongoing study of the thermal Casimir effect in 
the three-dimensional Ising universality class. Here we use the data generated
for various quantities to update the estimates of a number of universal 
amplitude ratios. Computing universal amplitude ratios, we follow the 
strategy of \cite{CaHa97,HaPi97}, where the spin-1/2 Ising model  had been
studied and more recently \cite{myXYApAm,myXYamplitude}, where we had studied 
an improved model in the XY universality class in three dimensions.

Our results are essentially consistent with previous Monte Carlo studies 
\cite{CaHa97,EnFrSe02,BlFe10} and the most recent analysis of high and 
low temperature series expansions \cite{pisaseries}. Typically we reduce 
the error bars by a factor of two to three compared with these studies.
Estimates obtained by using field theoretic methods are typically by a factor of ten
less precise than those obtained here. 

The outline of our paper is the following:
First we define the model and the observables that we have 
measured. Next we discuss the update algorithm and give details of our
simulations. Using the data obtained, we extract numerical estimates for various
universal amplitude ratios. These estimates are compared with those
obtained in previous Monte Carlo simulations, from high and low temperature 
series expansions, field theoretic methods and experiments. Finally we 
conclude.

\section{Model and observables}
The Blume-Capel model is characterized by the reduced Hamiltonian
\begin{equation}
\label{Blumeaction}
H = -\beta \sum_{<xy>} s_x s_y
  + D \sum_x s_x^2  - h \sum_x s_x \;\;  ,
\end{equation}
where the spin might assume the values $s_x \in \{-1,0,1 \}$. The 
sites on the simple cubic lattice are denoted by $x=(x_0,x_1,x_2)$ with 
$x_i =0, 1,...,L_i-1$. In the following we shall consider lattices with 
$L=L_0=L_1=L_2$ in the high temperature phase and $L_0=2 L$, $L_1=L_2=L$ 
lattices in the low temperature phase. Throughout we consider periodic 
boundary conditions.
The first sum in eq.~(\ref{Blumeaction}) 
runs over all pairs of nearest neighbor sites $<xy>$ on the lattice and
$\beta=1/k_B T$ is the inverse temperature. The partition function is given 
by
$Z = \sum_{\{s\}} \exp(- H)$  ,
where the sum runs over all spin configurations.
In the following we shall 
consider a vanishing external field $h=0$. The parameter $D$ controls the 
density of vacancies $s_x=0$. In the limit $D \rightarrow - \infty$ 
vacancies are completely suppressed and therefore the spin-1/2 Ising 
model is recovered.
In  $d\ge 2$  dimensions the model undergoes a continuous phase transition
for $-\infty \le  D   < D_{tri} $ at a $\beta_c$ that depends on $D$.
For $D > D_{tri}$ the model undergoes a first order phase transition.
Refs. \cite{des,HeBlo98,DeBl04}  give for the three-dimensional simple 
cubic lattice
$D_{tri} \approx 2.006, 2.05$ and $D_{tri} = 2.0313(4)$, 
respectively.

Numerically it has been shown that on the line of second order phase
transitions there is a value $D^*$ of the parameter $D$, 
where leading corrections to scaling vanish.
In ref. \cite{myhabil} we found $D^*=0.641(8)$. One should note that little
effort was made to estimate the systematical error due to subleading 
corrections to scaling.
Recently we have determined $D^*=0.656(20)$ \cite{mycritical},
where now systematical errors are taken into account.
In  \cite{mycritical} we have simulated the model at $D=0.641$ 
and $D=0.655$ in the neighborhood of the critical point. Using a standard
finite size scaling analysis we find
\begin{eqnarray}
\beta_c(0.641) & = &  0.38567122(5) \\
\beta_c(0.655) & = &  0.387721735(25)
\end{eqnarray}
as estimates of the inverse critical temperature. We also find that the 
amplitudes of leading corrections at $D=0.655$ are reduced by at least 
a factor of $30$ compared with the spin-1/2 Ising model.

\subsection{The energy density and the specific heat}
Here, we define the energy density as minus the derivative of the reduced 
free energy density with respect to $\beta$
\begin{equation}
\label{energy}
 E = \frac{1}{V} \frac{\partial}{\partial \beta}  \ln Z  = \frac{1}{V} \left \langle \sum_{<xy>}  s_x  s_y \right \rangle \;\;,
\end{equation}
where $V=L_0 L_1 L_2$.
The specific heat is the derivative of the  energy density with respect to 
$\beta$. One finds
\begin{equation}
 C  = \frac{\partial E}{\partial \beta} = \frac{1}{V} 
\left[\left \langle \left (\sum_{<xy>}  s_x  s_y \right)^2 \right \rangle
- \left \langle \sum_{<xy>}  s_x  s_y  \right \rangle^2 \right] \;\;.
\end{equation}

\subsection{The magnetic susceptibility and the second moment 
            correlation length in the high temperature phase}
The magnetic susceptibility $\chi$ and the second moment correlation length
$\xi_{2nd}$ are defined as
\begin{equation}
\chi  =  \frac{1}{V}
\left\langle \Big(\sum_x s_x \Big)^2 \right\rangle
\end{equation}
and
\begin{equation}
\label{xihigh}
\xi_{2nd}  = \sqrt{\frac{\chi/F-1}{4 \sin^2 \pi/L}} \;\;,
\end{equation}
where
\begin{equation}
F  =  \frac{1}{V} \left \langle
\Big|\sum_x \exp\left(i \frac{2 \pi x_k}{L} \right)
        s_x \Big|^2
\right \rangle
\end{equation}
is the Fourier transform of the correlation function at the lowest
non-zero momentum. In our simulations in the high temperature phase, 
we have measured $F$ for the three directions $k=0,1,2$ and have averaged
these three results.

\subsection{The magnetization, the
magnetic susceptibility and the correlation length in the low 
temperature phase}
\label{observables}
The magnetization in presence of a magnetic field is defined by
\begin{equation}
 m(h,L) = \frac{1}{V} \left \langle \sum_x s_x \right \rangle  \;\;,
\end{equation}
where we assume, for simplicity,  a fixed ratio $L_0/L$ with $L=L_1=L_2$.
The spontaneous magnetization is then defined as
\begin{equation}
m(0,\infty) = \lim_{h \searrow  0} \lim_{L \rightarrow \infty} m(h,L) \;\;,
\end{equation}
where first the thermodynamic limit is taken. In a Monte Carlo simulation
it is too cumbersome to follow this route. Note that $m(0,L)$ at a finite 
value of $L$ is however exactly zero for symmetry reasons.

To avoid this problem, Binder and Rauch \cite{BinderRauch} proposed the
following definition:
\begin{equation}
 m_{RMS}(0,L) = \frac{1}{V} \sqrt{ \left \langle \left(\sum_x s_x \right)^2 
 \right \rangle} \;\;.
\end{equation}

Here, following eqs.~(20,21) of \cite{Binderreview}, we use
\begin{equation}
\label{talapov}
m_{ABS}(0,L) = \frac{1}{V} \left \langle \left | \sum_x s_x \right | \right \rangle \;\;,
\end{equation}
which in the low temperature phase converges faster than $m_{RMS}(0,L)$.

The connected  two-point correlation function is given by
\begin{equation}
\label{basic}
 G(x,y) = \langle s_x s_y  \rangle - \langle s_x \rangle
                                          \langle s_y \rangle \;\;.
\end{equation}
In the low temperature phase, for $h=0$ we replace eq.~(\ref{basic}),
using eq.~(\ref{talapov}), by
\begin{equation}
\label{glow}
\left. G_{low}(x,y) \right |_{h=0}
  =\langle s_x s_y  \rangle -  m_{ABS}^2(0,L) \;\;.
\end{equation}

In order to project to zero-momentum states of the transfermatrix,
we consider the correlation function 
\begin{equation}
G(r) = \langle S_0 S_r \rangle - \langle S_0 \rangle \langle S_r \rangle \;\;.
\end{equation}
of time slices
\begin{equation}
 S_{x_0} = \frac{1}{\sqrt{L_1 L_2}} \sum_{x_1,x_2} s_{(x_0,x_1,x_2)} \;\;.
\end{equation}
Note that with this normalization, the correlation function has a
finite thermodynamic limit as $L_1,L_2 \rightarrow \infty$.
In the low temperature phase, for vanishing external field $h=0$ we replace
$\langle S_0 \rangle  \langle S_r \rangle$ by $L_1 L_2 m_{ABS}^2(0,L)$. 

The magnetic susceptibility can be written as
\begin{equation}
\label{chicorr}
\chi= \sum_{r=-\infty}^{\infty} G(r) \;\;.
\end{equation}

The effective correlation length is given by
\begin{equation}
\label{xieff}
\xi_{eff}(r) =-1/\ln\left(\frac{G(r+1)}{G(r)} \right) \;\;.
\end{equation}
The exponential correlation length is defined
as $\xi_{exp} = \lim_{r \rightarrow \infty} \xi_{eff}(r)$.
Since the transfermatrix is positive and
symmetric, $\xi_{eff}$ approaches $\xi_{exp}$ monotonically from below.
The second moment correlation length is defined by
\begin{equation}
\label{xi2nd}
 \xi_{2nd}^2 = \frac{\mu_2}{2 d \chi} \;\;,
\end{equation}
where $d=3$ is the dimension of the system and
\begin{equation}
\label{mu2}
\mu_2 = d \sum_{r=-\infty}^{\infty} r^2 G(r) \;\;.
\end{equation}
Note that in the thermodynamic limit, 
the definitions~(\ref{xihigh},\ref{xi2nd}) become equivalent.
In the low temperature phase we have computed $\chi$ and $\mu_2$ by
using eqs.~(\ref{chicorr},\ref{mu2}), respectively, in the 
following way: Up to a certain distance $R$ we have used $G(r)$ 
computed directly 
from the configurations that we have generated. Since the relative
statistical error increases exponentially with the distance $r$, for $r>R$ we 
have used instead
\begin{equation}
\tilde G(r) = G(R) \exp\left(-\frac{r-R}{\xi_{eff}(R)} \right) \;\;.
\end{equation}
In the following analysis we have used the data obtained by choosing  
$R \approx 4 \xi_{eff}(R)$.
We have checked that these results are consistent with those  
obtained for $R \approx 3 \xi_{eff}(R)$.

\section{The simulations}
\subsection{The Monte Carlo algorithm}
Analogous to \cite{BrTa}, we have simulated the Blume-Capel model
using a hybrid of local updates and single cluster updates \cite{Wolff}.
In the high temperature phase we have used as local update the heat-bath algorithm.
With the local update we run through the lattice in typewriter fashion.
Running through the lattice once is called one sweep in the following.
After two heat-bath sweeps we perform a certain number $N_{cl}$ 
of single cluster 
updates. We have chosen $N_{cl}$  to be roughly one third of the 
number of lattice sites $V$ divided by the average size of a 
cluster. In the following we shall denote two heat-bath sweeps followed 
by  $N_{cl}$ single cluster updates as one cycle of the update. In the 
high temperature phase we have used the cluster algorithm to compute 
improved estimators of the magnetic susceptibility and the second moment 
correlation length.

In the low temperature phase we have used a local Metropolis update
that is implemented in multispin coding technique \cite{multispin}. 
Details of our implementation can be found in \cite{mycritical}.
Here, after ten 
sweeps of the local update, we performed $N_{cl}$ single cluster updates.
Also here we have chosen $N_{cl}$  to be roughly one third of the
number of lattice sites $V$ divided by the average size of a
cluster. Here we denote ten sweeps of the local update followed by
$N_{cl}$ single cluster updates as one cycle. In the low temperature 
phase we did not use cluster improved estimators since they 
do not reduce the statistical error significantly in this phase. 

As random number generator we have used the  SIMD-oriented Fast
Mersenne Twister algorithm \cite{twister}.

\subsection{Simulations in the high temperature phase}
\label{hightemp}
First we have checked which lattice sizes are needed to keep the deviation 
from the thermodynamic limit smaller than the statistical error of the
observables that we measure. Based on finite size scaling theory 
\cite{Barber},  
we expect that the $L$-dependence of a singular observable $A$ is given by
\begin{equation}
 A(L,\beta) = A(\infty,\beta) \times [1 + g_A(L/\xi(\beta))]  \;\;,
\end{equation}
where we have ignored corrections to scaling.  In the absence of a massless
mode, as it is the case here, 
\begin{equation}
 g_A(L/\xi) \simeq c_A \exp(-L/\xi) 
\end{equation}
for large values of $L/\xi$. At $D=0.655$ and $\beta=0.372$ we have simulated
lattices with the linear lattice sizes $L=18$, $20$, $22$, $24$, $26$, $28$,
$32$, $40$ and $48$ to check the size dependence of the observables. In total
these simulations took about 6 days of CPU time on single core of a Quad-Core
Opteron(tm) 2378 CPU (2.4 GHz). Our results are summarized in table
\ref{finitesizelow}. We give the number of update cycles (stat), 
the number of single cluster updates $N_{cl}$ per update cycle and the 
estimates of the energy density $E$, the magnetic susceptibility $\chi$
and the second moment correlation length $\xi_{2nd}$. We have fitted these
results with the ansatz $A(L) = A(\infty) + c_A \exp(-L/\xi_{exp})$, where we 
have taken $\xi_{exp}=3.09394(13)$, which is the result for the exponential 
correlation length that we have obtained for $L=48$. Skipping the data for
$L=18$ we get an acceptable $\chi^2/$d.o.f. for all three quantities. The 
result for the thermodynamic limit $A(\infty)$ is given in the last row 
of table \ref{finitesizelow}. The correction amplitudes are $c_E=0.114(3)$, 
$c_{\chi}=-21.2(1.1)$ and $c_{\xi_{2nd}}=-1.28(8)$. This means that the 
deviation from the thermodynamic limit is of the same size as the statistical
error that we have reached here for $L/\xi  \approx  10$  for all three
quantities that we have studied. 

In the following simulations we have chosen $L \gtrapprox 10 \xi_{2nd}$
throughout.
Since for most of our simulations $L$ is clearly larger than $10 \xi_{2nd}$
and the relative statistical error of the magnetic susceptibility and the 
second moment correlation length is larger than that of the results
discussed above, we
expect that deviations from the thermodynamic limit can be safely ignored. 

\begin{table}
\caption{\sl \label{finitesizelow} The energy density $E$, the magnetic 
susceptibility $\chi$ and the second moment correlation length $\xi_{2nd}$
at $D=0.655$ and $\beta=0.372$ for various linear lattice sizes $L$.
Furthermore we give the number of update cycles (stat) and the number of 
single cluster updates $N_{cl}$ per update cycle. In the last raw
we give the result of our extrapolation to the thermodynamic limit, as
discussed in the text.
}
\begin{center}
\begin{tabular}{lcrlll}
\hline
 \multicolumn{1}{c}{$L$}
 & \multicolumn{1}{c}{stat$/10^6$}
 &\multicolumn{1}{c}{$N_{cl}$}
 & \multicolumn{1}{c}{$E$}
 & \multicolumn{1}{c}{$\chi$}
 & \multicolumn{1}{c}{$\xi_{2nd}$} \\
\hline
18 & 50\phantom{.0} & 200 & 0.492029(5) &  24.4851(18)  & 3.07593(13)\\ 
20 & 40\phantom{.0} & 300 & 0.491812(5) &  24.5198(17)  & 3.07794(13)\\ 
22 &35\phantom{.0} & 400 & 0.491721(4) &  24.5405(16)  & 3.07910(12)\\ 
24 &30.9 & 500&0.491678(4)&  24.5469(15)  & 3.07945(11)\\ 
26&32.4 & 700 & 0.491658(3)& 24.5502(13)  & 3.07963(10)\\ 
28 & 20\phantom{.0} &1000 & 0.491643(4) &  24.5532(14)  & 3.07985(11)\\ 
32 &10\phantom{.0} &1000 & 0.491635(5) &  24.5538(18)  & 3.08000(13)\\ 
40&\phantom{0}7.5 &2500 & 0.491633(4) &  24.5532(14)  & 3.07994(11)\\ 
48&\phantom{0}5\phantom{.0} &2000 & 0.491633(4) &  24.5545(15)  & 3.08008(12)\\
\hline
 $\infty$ & & & 0.4916314(17) & 24.5549(7) & 3.08000(5) \\
\hline
\end{tabular}
\end{center}
\end{table}

In the case of $D=0.655$ we have simulated at 201 different
values of $\beta$ starting from $\beta=0.25$ up to $\beta=0.3872$. 
For each value of $\beta$ we have performed 500000 update cycles.
At $\beta=0.3872$ we have simulated an $L=300$ lattice and find $\xi_{2nd} = 26.698(7)$.
In the case of $D=0.641$ we have only simulated at $12$ different 
values of $\beta$ from $\beta = 0.3827$, where $\xi_{2nd}=8.8993(7)$, 
up to $\beta = 0.3849$, where $\xi_{2nd}=20.859(3)$.  Here, for the two smallest
values of $\beta$ we performed about $3 \times 10^6$ update cycles and about 
$1.5 \times 10^6$ for the larger ones. 
The simulations in the high temperature phase at $D=0.655$ and $0.641$ together
took about one year of CPU time on a 
single core of a  Quad-Core Opteron(tm) 2378 CPU (2.4 GHz).

\subsection{Simulations in the low temperature phase}
\label{lowtemp}
Also here we have checked which lattice sizes are needed to keep deviations
from the thermodynamic limit sufficiently small to be safely ignored.
To this end, we performed simulations at $D=0.655$ and $\beta=0.405$ 
using the linear lattice sizes $L=12$, $16$, $20$, $24$, $30$ and $40$.
These simulations took about three weeks  of CPU time on a
single core of a  Quad-Core Opteron(tm) 2378 CPU (2.4 GHz).
Our results for the 
various observables are summarized in table \ref{finitesizelow}. Starting 
from $L=20$ the results are consistent among each other.

Below we followed the recommendation of \cite{CaHa97} and have used lattices
with $L > 20 \xi_{2nd}$. Given the observations made here, this is surely
a safe choice.

\begin{table}
\caption{\sl \label{finitesizelow} Estimates for the energy density $E$, 
the magnetization $m$, the magnetic susceptibility $\chi$, the second 
moment correlation length $\xi_{2nd}$ and the exponential correlation 
length $\xi_{exp}$ at $D=0.655$ and $\beta=0.405$ for various linear 
lattice sizes $L$. The results for $\chi$ and $\xi_{2nd}$ 
are obtained with $R=4$ and correspondingly, 
$\xi_{exp}$ is approximated by $\xi_{eff}(6)$, since $4 \xi_{exp} \approx 6$. 
Furthermore we give the 
number of update cycles (stat) that we have performed. This number 
includes the factor $64$ of copies of the system that we have simulated.
In all cases we have performed two single cluster updates for each update
cycle.
}
\begin{center}
\begin{tabular}{lclllll}
\hline
 \multicolumn{1}{c}{$L$}
 & \multicolumn{1}{c}{stat$/10^6$}
 & \multicolumn{1}{c}{$E$}
 & \multicolumn{1}{c}{$m$}
 & \multicolumn{1}{c}{$\chi$}
 & \multicolumn{1}{c}{$\xi_{2nd}$}
 & \multicolumn{1}{c}{$\xi_{exp}$} \\
\hline
12 &640&1.063260(12)&0.514811(5)&4.4891(4)&1.5483(3)&1.6056(6)  \\
16 &320&1.063467(11)&0.514997(5)&4.4317(4)&1.5271(3)&1.5778(7)  \\
20 &320&1.063487(8) &0.515010(3)&4.4281(4)&1.5258(3)&1.5759(6)  \\
24 &190&1.063474(8) &0.515005(3)&4.4283(4)&1.5259(4)&1.5761(8) \\
30 &130&1.063482(7) &0.515008(3)&4.4284(4)&1.5258(4)&1.5762(8)  \\
40 &\phantom{0}32&1.063474(9) &0.515004(4)&4.4277(7)&1.5243(7)&1.5729(15) \\ 
\hline
\end{tabular}
\end{center}
\end{table}
In the case of $D=0.655$ we have simulated 64 different values of $\beta$
starting from $\beta=0.3884$, where $\xi_{2nd}=11.687(45)$ up to 
$\beta=0.42$, where $\xi_{2nd}=1.0293(7)$. In addition we have simulated
at 85 values of $\beta$  up to $\beta=0.60$, 
where we have only measured the energy density.
In the case of $D=0.641$ we have simulated at 12 different values of $\beta$
starting from $\beta=0.3866$, where $\xi_{2nd}=9.589(19)$, up to 
$\beta=0.3899$, where $\xi_{2nd}= 4.3749(34)$.
In total these simulations in the low temperature phase took about 10 years
of CPU time on a single core of a  Quad-Core Opteron(tm) 2378
CPU (2.4 GHz).

\section{Universal Amplitude Ratios}
First we have computed the ratio of the amplitudes of the magnetic 
susceptibility in the high and the low temperature phase. To this 
end one could fit the data for the magnetic susceptibility  with an 
ansatz such as eq.~(\ref{chicorrection}) for the data in the high and the 
low temperature phase separately. Using the results for $C_+$ and $C_-$
obtained this way one could compute the ratio $C_+/C_-$. Instead, 
following ref. \cite{CaHa97} we use a different strategy.
The amplitude ratio can be defined as
\begin{equation}
\frac{C_+}{C_-} = \lim_{t \searrow 0} \frac{\chi(t)}{\chi(-t)} \;\;.
\end{equation}
Following this definition, we have first calculated the the ratio
$\chi(t)/\chi(-t)$ at finite  values of $t=\beta_c-\beta$. To this end, 
we have computed for the values $\beta_{low}$, where we have simulated 
in the low temperature phase corresponding values 
$\beta_{high}= 2 \beta_c-\beta_{low}$.  Here we made no effort  to simulate
exactly at these values of $\beta_{high}$. Instead we interpolate  between 
the values that we have simulated. To this end, we take the 
$\beta_{-} \le \beta_{high} \le \beta_{+}$ which are  closest to 
$ \beta_{high}$.  We compute 
$c(\beta_{\pm}) = \chi(\beta_{\pm}) (\beta_c - \beta_{\pm})^{\gamma}$, using 
$\gamma=1.23719$ \cite{mycritical}. 
Then we linearly interpolate to get an estimate of 
$c(\beta_{high})$.  Finally we compute
$\chi(\beta_{high}) = c(\beta_{high}) (\beta_c - \beta_{high})^{-\gamma}$.
Since we have simulated a large number of $\beta$-values, the systematical
error introduced by this interpolation should be negligible.
\begin{figure}
\begin{center}
\includegraphics[width=14.5cm]{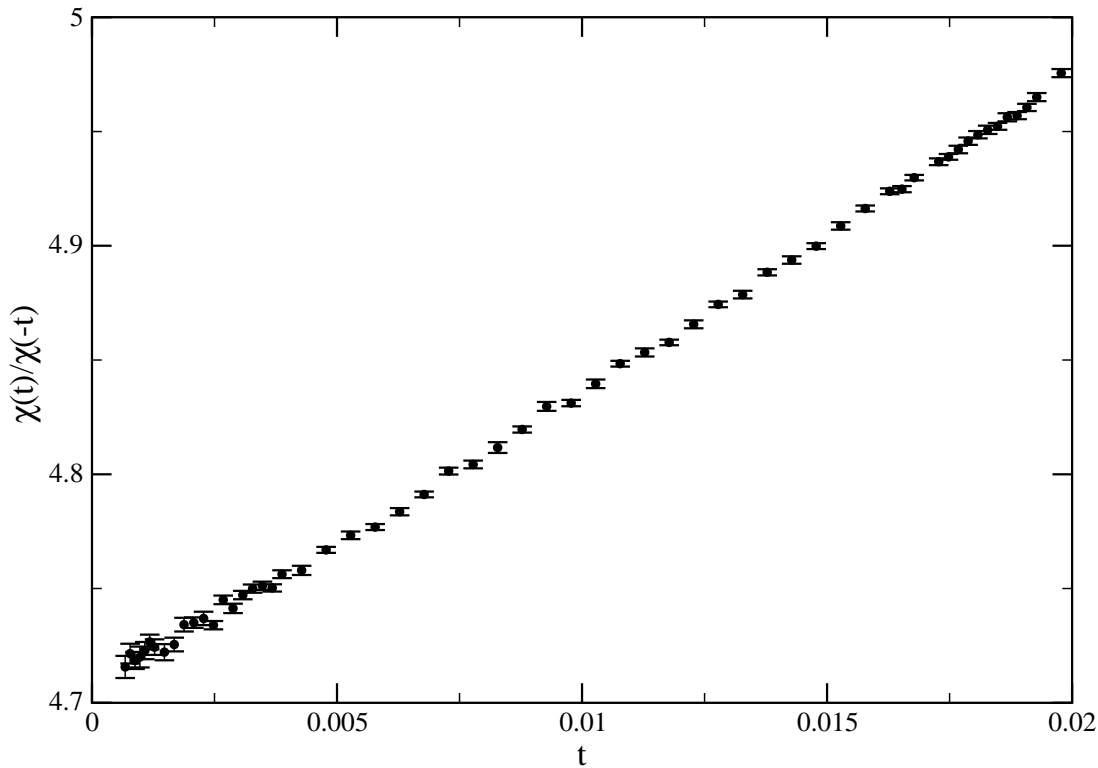}
\caption{\label{chiplot}
We plot the ratio $\chi(t)/\chi(-t)$ as a function of $t=\beta_c -\beta$
computed from our data for $D=0.655$. For a discussion see the text.
}
\end{center}
\end{figure}
In figure \ref{chiplot} we have plotted our results for $\chi(t)/\chi(-t)$ as 
a function of $t$.  From RG theory we expect that 
\begin{equation}
\label{ratiocorrection}
\frac{\chi(t)}{\chi(-t)}=\frac{C_+}{C_-} + a t^{\theta} +  b t 
                      + c t^{\theta'} + d t^{2 \theta} + e t^{\gamma} + ...\;\;,
\end{equation}
where $a$ and in particular $d$ should be small here, since  $D=0.655$ is 
a good approximation of $D^*$.  Following ref. \cite{NewmanRiedel} 
$\theta'=1.05(7)$. Therefore the term $ c t^{\theta'} $ can be hardly 
discriminated from the analytic correction $ b t$.  The term $e t^{\gamma}$
is caused by the analytic background of the magnetic susceptibility. 
Figure  \ref{chiplot}  suggests that $\chi(t)/\chi(-t)$ is essentially
a linear function of $t$. Therefore, the terms explicitly given in 
eq.~(\ref{ratiocorrection}) should be sufficient to fit $\chi(t)/\chi(-t)$. 
In particular we have fitted our data with the ans\"atze
\begin{eqnarray}
 \frac{\chi(t)}{\chi(-t)}&=& \frac{C_+}{C_-} +  b t  \label{chifit1} \\
 \frac{\chi(t)}{\chi(-t)}&=& \frac{C_+}{C_-} + a t^{\theta} + b t 
\label{chifit2}  \\
\frac{\chi(t)}{\chi(-t)}&=& \frac{C_+}{C_-}  + a t^{\theta} + b t  
+ e t^{\gamma} \label{chifit3}
\end{eqnarray}
using $\theta=0.524$ and $\gamma=1.23719$. Using these ans\"atze we have 
performed a large number of fits. Below we give the results of those fits
that include a maximal number of data points under the condition that 
$\chi^2/$d.o.f. is close to one. The fits are done using our data 
for $D=0.655$ if not stated otherwise.
Fitting the data that satisfy $t < 0.005$ with the ansatz~(\ref{chifit1})
we get $C_+/C_- = 4.7089(14)$, $b=12.0(4)$ and $\chi^2/$d.o.f.$=24.6/20$. 
Fitting all available data for $D=0.641$ with ansatz~(\ref{chifit1})
we get $C_+/C_- = 4.7145(21)$, $b=11.4(9)$ and $\chi^2/$d.o.f.$=9.6/8$.
Using ansatz~(\ref{chifit2}), fitting data with $t < 0.019$  we get
$C_+/C_- = 4.718(2)$, $a=-0.43(6)$, $b=15.6(3)$ and $\chi^2/$d.o.f.$=55.8/53$.
Fitting the data that satisfy $t < 0.022$ with the ansatz~(\ref{chifit3}) we
get $C_+/C_- = 4.712(6)$, $a=-0.05(30)$, $b=8.7(4.9)$, $e=12.3(8.2)$ and
$\chi^2/$d.o.f.$=66.6/57$.
As our final result we quote
\begin{equation}
  \frac{C_+}{C_-} = 4.713(7)  \;\;,
\end{equation}
which is chosen such that it covers all results, including their error bars,
of the fits quoted above. We have estimated the error due to the uncertainty
of $\beta_c$ by redoing some of the fits using ratios computed
with $\beta_{high} = 2 (\beta_c + \mbox{error})-\beta_{low}$. 
We find that it is  clearly smaller than the error quoted above.

Next we have computed the amplitude ratio
\begin{equation}
\frac{f_{2nd,+}}{f_{2nd,-}} = \lim_{t \searrow 0} \frac{\xi_{2nd}(t)}{\xi_{2nd}(-t)} \;\;.
\end{equation}
The calculation is analogous to that of the ratio $C_+/C_-$. Therefore 
we abstain from giving details and directly quote our final result
\begin{equation}
\frac{f_{2nd,+}}{f_{2nd,-}} = 1.939(5) \;\;.
\end{equation}

Next we have computed the RG-invariant quantity 
\begin{equation}
 Q_c  = \frac{f_{2nd,+}^3 B^2}{C_+} \;\;,
\end{equation}
where $B$ is the amplitude of the spontaneous magnetization in the 
low temperature phase.
To this end, we have first evaluated $r=\xi_{2nd}^3/\chi$ for all values 
of $\beta$ that we have simulated in the high temperature phase, where
we have computed the statistical error of this combined 
quantity using the Jackknife method.  Then we have computed 
\begin{equation}
Q_c = \lim_{t \searrow 0}  r(t) m^2(-t) 
\end{equation}
in the same fashion as we have computed $C_+/C_-$ and $f_{2nd,+}/f_{2nd,-}$ 
above.
As our final estimate we quote
\begin{equation}
Q_c = 0.3293(2)  \;\;.
\end{equation}

\begin{figure}
\begin{center}
\includegraphics[width=14.5cm]{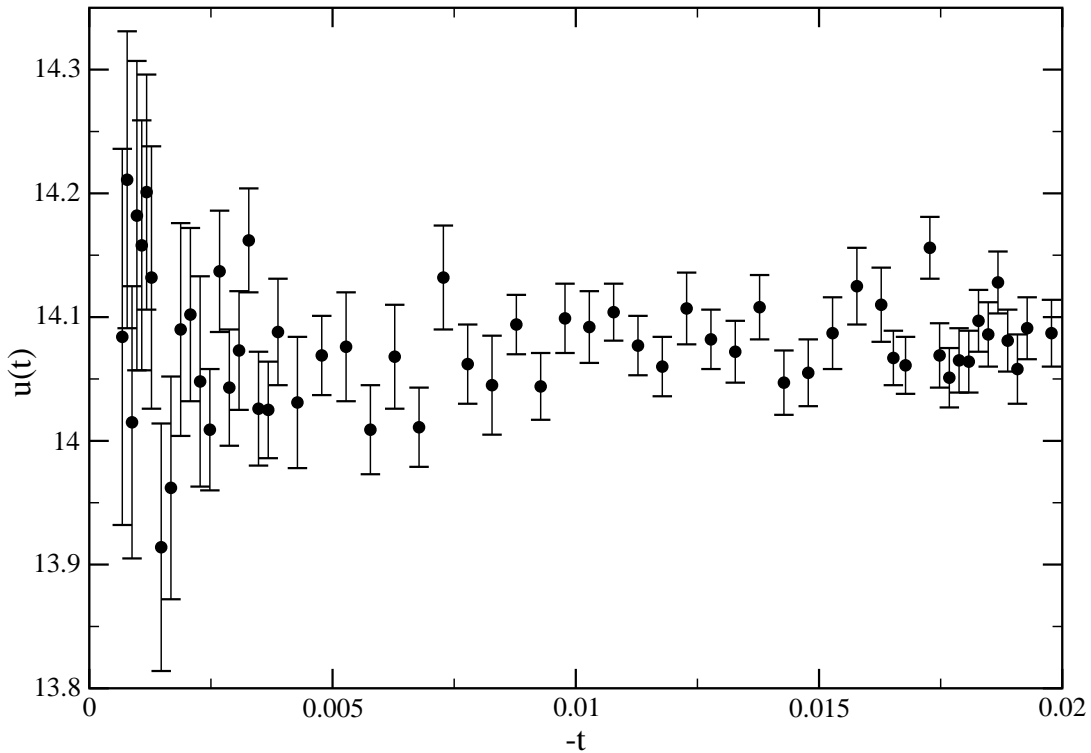}
\caption{\label{Uplot}
We plot the ratio $u(t)=3 \chi(t)/\xi_{2nd}^3(t) m^2(t)$ as a function of
$-t=\beta-\beta_c$ computed from our data for $D=0.655$. For a discussion 
see the text.
}
\end{center}
\end{figure}

The renormalized coupling in the low temperature phase is given by
\begin{equation}
 u^* = \frac{3 C_-}{f_{2nd,-}^3 B^2} =
 \lim_{t \nearrow 0} \frac{3 \chi(t)}{\xi_{2nd}^3(t) m^2(t)} \;\;.
\end{equation}
In figure \ref{Uplot} we plot $3 \chi(t)/\xi_{2nd}^3(t) m^2(t)$  
as a function of $-t$. 
Since only quantities in the low temperature phase are involved, 
there should be no analytic correction. However, since 
$\theta' \approx 1$ we kept a term $b t$ in our ans\"atze. Based on 
various fits we arrive at the final estimate
\begin{equation}
 u^* = 14.08(5)  \;\;.
\end{equation}

Now let us consider the ratio $\xi_{exp}/\xi_{2nd}$.  It turns out 
that it is difficult to determine the exponential correlation length 
accurately in the low temperature have. The time slice correlation 
function behaves as 
\begin{equation}
\label{multiexp}
 G(r) = c_1 \exp(-r/\xi_1) + c_2 \exp(-r/\xi_2) + ... \;\;.
\end{equation}
Since the ratio $\xi_1/\xi_2 = 1.83(3)$ \cite{ourstuff} is rather small, 
the effective correlation length $\xi_{eff}$, eq.~(\ref{xieff}), converges
only rather slowly to $\xi_{exp}$.  On the other hand, the relative 
statistical error of $G(r)$ increases exponentially. Therefore very 
large distances that are needed to get a small deviation of $\xi_{eff}$ from
$\xi_{exp}$ are not accessible. As compromise, we have taken $\xi_{eff}(R)$
with $R \approx 4 \xi_{eff}(R)$ as our final estimate. To check the 
systematical error introduced this way, we have compared our result with 
that for $R \approx 3 \xi_{eff}(R)$.  As our final estimate we quote 
\begin{equation}
\frac{f_{exp,-}}{f_{2nd,-}} = 1.020(5) \;\;.
\end{equation}
Here the error should cover both the systematical deviation of 
$\xi_{eff}$ from $\xi_{exp}$ as well as systematical errors due to 
subleading corrections that are not included in our fits. 
In order to get more precise results for $\xi_{exp}$ and as a 
consequence for $f_{exp,-}/f_{2nd,-}$ a variational analysis of a large
set of correlation functions, as it has been done in ref. \cite{ourstuff} 
would be useful. Furthermore the method of \cite{LuWe} to reduce 
the variance of correlation functions could help to compute $G(r)$ accurately
at large distances $r$. 

\subsection{Ratios that involve the specific heat}
In order to compute amplitude ratios that involve the specific heat, we
have analyzed our data for the energy density which can be accurately 
determined in the simulation. In the case of the energy density we have 
to separate the analytic background and the singular part, which is needed 
here. In the neighborhood of the critical point, the energy density behaves
as 
\begin{equation}
 E = E_b + E_s \;\;,
\end{equation}
where the analytic background can be Taylor expanded around the 
critical point:
\begin{equation}
E_b(\beta)=E_{ns} + C_{ns} (\beta-\beta_c) + d_{ns} (\beta-\beta_c)^2
            + ... \;\;.
\end{equation}
The singular part is given by
\begin{equation}
 E_{s} =  a_{\pm}  |t|^{-\alpha} \times (1 + b_{\pm} |t|^{\theta} + c t + ...) \;\;.
\end{equation} 

In a first step we have analyzed data generated in relation with 
\cite{mycritical} for cubic systems with a linear size up to $L=360$ and periodic boundary 
conditions directly at the critical point.
At the critical point the energy density behaves as
\begin{equation}
 E = E_{ns}  + a L^{-3+1/\nu} (1 +  c L^{-\omega} +  ...)
\end{equation}
and the specific heat as
\begin{equation}
 C = C_{ns}  + b L^{-3+2/\nu} (1 +  d L^{-\omega} +  ...) \;\;.
\end{equation}
Here we perform fits fixing $\nu=0.63002(10)$ as obtained in \cite{mycritical}.
Our final estimate is taken
from fits without any correction term and all lattice sizes that are larger or equal to
$L_{min}=24$ taken into account. Systematic errors are estimated by performing
fits that include corrections with an exponent that is either $0.832$, $1.6$
or $2$. For $D=0.655$ we get
\begin{equation}
 E_{ns} = 0.602111(1) +
 0.006 \times (\nu - 0.63002) +42 \times (\beta_c - 0.387721735)
\end{equation}
and 
\begin{equation}
a = 1.7490(5) + 14 \times (\nu-0.63002) -1800 \times (\beta_c - 0.387721735) \;\;.
\end{equation}
We have redone the fits with slightly shifted values of the input parameters
$\nu$ and $\beta_c$ to obtain the dependence of $E_{ns}$ and $a$ on these parameters. 
For the specific heat at $D=0.655$ we get
\begin{equation}
C_{ns} = -19.1(1) -1700 \times (\nu-0.63002) -1300000 \times (\beta_c-0.387721735)
\end{equation}
and
\begin{equation}
 b = 25.30(5) +1350 \times (\nu-0.63002) + 620000 \times (\beta_c-0.387721735)
\end{equation}
from fits with $L_{min}=64$.  The error is dominated by systematical errors
that we have estimated from fits that include corrections to scaling.

For the energy density at $D=0.641$ we get
\begin{equation}
 E_{ns} = 0.604870(2) +
 0.01 \times (\nu - 0.63002) + 41 \times (\beta_c - 0.38567122)
\end{equation}
and
\begin{equation}
a = 1.749(1) + 14 \times (\nu-0.63002) - 1800 \times (\beta_c - 0.38567122) \;\;.
\end{equation}
For the specific heat at $D=0.641$ we get
\begin{equation}
C_{ns} = -19.1(2) - 1700  \times (\nu-0.63002) - 1000000 \times (\beta_c-0.38567122)
\end{equation}
and 
\begin{equation}
 b = 25.3(1) + 1350 \times (\nu-0.63002) + 500000  \times (\beta_c-0.38567122) \;\;.
\end{equation}

Next we have analyzed our data for the thermodynamic limit in the neighborhood 
of the critical point using the ansatz
\begin{equation}
\label{critical2}
 E(\beta) = E_{ns} + C_{ns} (\beta-\beta_c)
             + a_{\pm} |\beta-\beta_c|^{1-\alpha}
             + d_{ns} (\beta-\beta_c)^2
             + b_{\pm} |\beta-\beta_c|^{2-\alpha} \;\;,
\end{equation}
where $E_{ns}$, $C_{ns}$ obtained above and 
$\beta_c=0.387721735(25)$ and $\alpha=0.10994(30)$ \cite{mycritical} are input parameters,
while $a_{\pm}$, $d_{ns}$ and $b_{\pm}$ are the 5 free parameters of the fit. 
Using the results of these fits we have computed $A_+/A_-=-a_+/a_-$ and 
$P=(1-A_+/A_-)/\alpha$, which depends less on the input value for $\alpha$ than $A_+/A_-$. 

Fitting all data for $D=0.655$ in the interval $[\beta_c-0.004,\beta_c+0.004]$
we get $A_+/A_-=0.53611(7)$, $P=4.2195(6)$ and $\chi^2/$d.o.f. $= 55.5/51$ 
using the central values of the input parameters.  
For the interval $[\beta_c-0.0075,\beta_c+0.0075]$ we get
$A_+/A_-=0.53614(3)$, $P=4.2192(3)$ and $\chi^2/$d.o.f. $=165.7/93$. As 
a check we have also fitted with an ansatz, where we have added a term 
$\propto (\beta-\beta_c)^3$ compared with the ansatz~(\ref{critical2}). 
The results for $A_+/A_-$ and $P$ change little compared with those
given above. It turns out that the error of $A_+/A_-$ and $P$ is actually
dominated by the error induced by the uncertainty of our input parameters, 
$E_{ns}$, $c_{ns}$, $\beta_c$ and $\alpha$.
In order to estimate this error, we have repeated
the fits using shifted values of these input parameters. For example,
we have replaced $E_{ns}$ by $(E_{ns}+\mbox{error})$. 

In order to check for the effect of leading corrections to scaling, 
we have fitted all our data at $D=0.641$ using the ansatz~(\ref{critical2}).
We find   $A_+/A_-=0.53624(11)$, $P=4.2183(10)$ and 
$\chi^2/$d.o.f. $=22.3/19$ using the central values of the input parameters. 
This means that the results obtained at $D=0.641$ and $D=0.655$ are 
fully consistent.

We arrive at the final estimates
\begin{equation}
\frac{A_+}{A_-}=0.536(2)  \;\;,
\end{equation}
where the error is dominated by the uncertainty of $\alpha$, followed 
by the uncertainty of $C_{ns}$. In contrast 
\begin{equation}
P=4.22(1) 
\end{equation}
depends much less on the value of $\alpha$. 
Its error is dominated by the uncertainty of $C_{ns}$. This different
behavior of $A_+/A_-$ and $P$ is actually much more important in the
case of the XY-universality class, where $\alpha$ is close to zero and 
therefore the relative accuracy of $\alpha$ is much smaller than in the 
present case.

In order to compute the quantities
\begin{equation}
 Q_{\xi,+} =   \alpha A_+ f_{2nd,+}^3
\end{equation}
and 
\begin{equation}
 Q_{\xi,-} =   \alpha A_- f_{2nd,-}^3
\end{equation}
we have approximated the singular part of the energy density by
\begin{equation}
 E_s(t) = E(t) -E_{ns} - C_{ns} t \;\;.
\end{equation}
Then we have computed
\begin{equation}
q(t) = t \xi_{2nd}^3(t)  E_s(t) \;\;.
\end{equation}
The quantity  $Q_{\pm}$ is then given by
\begin{eqnarray}
 Q_+ &=& \alpha (1-\alpha) \lim_{t \searrow 0} q(t)  \\
 Q_- &=& \alpha (1-\alpha) \lim_{t \nearrow 0} q(t)  \;\;.
\end{eqnarray}

We have fitted our data with the ansatz  
\begin{equation}
 q(t) = q^* + a t \;\;.
\end{equation}
In the high temperature phase we find by fitting all data with $t < 0.006$ 
the result $q^*= 0.19412(3)$. We have redone this analysis with shifted
values of $C_{ns}$ and $E_{ns}$ to estimate the effect on our result
for $q^*$. 
It turns out that the error is dominated by the errors induced by the 
uncertainty of $C_{ns}$ and $E_{ns}$.
We have also redone the analysis using our data for $D=0.641$.
We get an estimate for $q^*$ that is fully consistent with that for $D=0.655$.
We arrive at the final result
\begin{equation}
 Q_{+} = \alpha (1-\alpha) q^* = 0.01899(10) \;\;.
\end{equation} 

Performing a similar analysis we arrive at 
\begin{equation}
Q_{-} = 0.00487(2)  \;\;.
\end{equation}

\section{Comparison with results given in the literature}
\subsection{Monte Carlo simulations and high and 
low temperature series}
In table \ref{comparelattice}
we confront our results with those of previous Monte Carlo simulations
\cite{CaHa97,HaPi97,EnFrSe02,BlFe10}, with a comprehensive analysis 
of high and low temperature series \cite{pisaseries} and the low temperature
series estimate of $u^*$ given in \cite{PeVi98}. In \cite{pisaseries} 
a parametric representation of the equation of state has been used 
to obtain results for the critical isotherm and the low temperature phase
from high temperature series.   For an 
exhaustive overview of the literature see table 11 of \cite{PeVi02}. 

In \cite{CaHa97,HaPi97} we have simulated the spin-1/2 Ising model on simple 
cubic lattices of a linear size up to $L=120$ and $L=128$, respectively.
Also the authors of \cite{BlFe10} have simulated the spin-1/2 Ising model on 
the simple cubic lattice. They have simulated 
at a large number of $\beta$-values in both phases of the model on lattices 
of a size  up to $L=128$.
In \cite{EnFrSe02} the $\phi^4$ model on the simple cubic lattice has 
been simulated at $\lambda=1.1$, which is the estimate of $\lambda^*$ 
obtained in ref. \cite{myphi4}. The authors have simulated lattices up 
to the size $L=120$.  In addition to simulations at a vanishing external
field $h=0$, they have simulated $h\ne 0$ at the critical 
temperature. This allowed them to compute additional universal amplitudes
ratios that we do not discuss here.

\begin{table}
\caption{\sl \label{comparelattice} Results for universal amplitude ratios
obtained by high and low temperature series expansions of three different 
improved lattice models \cite{pisaseries} and Monte Carlo simulations of
the spin-1/2 Ising model \cite{CaHa97,HaPi97,BlFe10} and the improved 
$\phi^4$ model \cite{EnFrSe02}. In all these cases a simple cubic lattice 
has been studied.
}
\begin{center}
\begin{tabular}{lllllllll}
\hline
 \multicolumn{1}{c}{Ref.}
 & \multicolumn{1}{c}{$A_+/A_-$}
 & \multicolumn{1}{c}{$C_+/C_-$}
 & \multicolumn{1}{c}{$\frac{f_{2nd,+}}{f_{2nd,-}}$}
 & \multicolumn{1}{c}{$\frac{f_{exp,-}}{f_{2nd,-}}$}
 & \multicolumn{1}{c}{$Q_+$}
 & \multicolumn{1}{c}{$Q_-$}
 & \multicolumn{1}{c}{$u^*$}
 & \multicolumn{1}{c}{$Q_c$} \\
\hline
here & 0.536(2) & 4.713(7)  & 1.939(5) &1.020(5) & 0.01899(10) & 0.00487(2) &  14.08(5) & 0.3293(2) \\
\cite{PeVi98}    & &        &         &          & & &14.25(12)&    \\
\cite{pisaseries} & 0.532(3) & 4.76(2) & 1.956(7)& & 0.01880(8) & 0.00472(5) & & 0.3315(10)\\
\cite{CaHa97} & & 4.75(3)& 1.95(2) & 1.017(7) & & &14.3(1) & 0.328(5) \\
\cite{HaPi97} &0.560(10)&&         &          & & &         &         \\
\cite{EnFrSe02}& & 4.756(28)& 1.935(14)&      & & &         & 0.326(3) \\
\cite{BlFe10}&0.532(7)&     &          &      & & &         &          \\
\hline
\end{tabular}
\end{center}
\end{table}

Essentially our results confirm those of the previous work. Even in the worst 
case, the deviation between our result and that of the other works summarized 
in table \ref{comparelattice} is less than three times the combined error.

\subsection{Field theoretic methods}
In table \ref{comparefield} we have summarized results obtained from the 
$\epsilon$-expansion and perturbation theory in three dimensions fixed.
Mostly we have taken these results from table 12 of ref. \cite{PeVi02}. 
Note that in ref. \cite{GuZi98} the field theoretic methods have been used 
in connection with a parametric representation of the equation of state.
In ref. \cite{Bagnuls} $A_+ C_+/B^2 = 0.0594(11)$ is given. By using  the 
value of $Q_+$ given e.g. by  \cite{Bagnuls85}  $Q_c$ can be computed.
Here we only 
report those amplitude ratios that we have computed in this work. 
For a comprehensive list of amplitude ratios see table 12 
of ref. \cite{PeVi02}. Essentially the field theoretic results are consistent 
ours, albeit their accuracy is clearly lower than ours.  The errors
for $Q_+$ given by \cite{BeGo80,Bagnuls85} seems to be underestimated.

\begin{table}
\caption{\sl \label{comparefield} Results for universal amplitude ratios 
obtained by using the $\epsilon$-expansion ($\epsilon$)
and  perturbation theory in three dimensions (3D) fixed.
For the definition of the amplitude ratios and a discussion see the text.
}
\begin{center}
\begin{tabular}{lcllllll}
\hline
 \multicolumn{1}{c}{Ref.}
 &  \multicolumn{1}{c}{Method}
 & \multicolumn{1}{c}{$A_+/A_-$}
 & \multicolumn{1}{c}{$C_+/C_-$}
 & \multicolumn{1}{c}{$\frac{f_{2nd,+}}{f_{2nd,-}}$}
 & \multicolumn{1}{c}{$Q_+$}
 & \multicolumn{1}{c}{$u^*$}
 & \multicolumn{1}{c}{$Q_c$} \\
\hline
\cite{BrGuZi74} & $\epsilon$&          &   4.8    &1.91&  &    &  \\   
\cite{AhHo76} & $\epsilon$ &0.55       &   4.8    &    &  &    &  \\
\cite{BeGo80}& $\epsilon$ &       &          &    &0.01966(17)  &  &  \\
\cite{NiAl85}& $\epsilon$ & 0.44       &   4.9    & &0.0223    &    &  \\
\cite{Ber86} & $\epsilon$ & 0.524(10) &          &    &  &    &  \\
\cite{GuZi98} & $\epsilon$ & 0.527(37)& 4.73(16) &  &    &    &  \\
\cite{Bagnuls} & 3D  & 0.541(14) & 4.77(30) &      &   &      &  0.331(9) \\
\cite{Bagnuls85} & 3D   &           &         &  & 0.01968(15)  & & \\
\cite{Dohmetal} & 3D  & 0.540(11) &          &  &  &    &    \\
\cite{GuKuMu96} & 3D  &           & 4.72(17) &2.013(28) &    & 14.2 &  \\
\cite{GuZi98} & 3D    & 0.537(19) & 4.79(10) &  &  &  &      \\
\cite{DoSt03} & 3D  &           &                   &  & 0.0203  & & \\
\hline 
\end{tabular} 
\end{center} 
\end{table}

\subsection{Experiments}
Here we just mention the results of two experimental works to give the reader an idea 
of the accuracy that can be reached. Studying a mixture of succinonitrile and water 
the authors of \cite{No01} found $A_+/A_- = 0.536 \pm 0.005$ and $Q_+= 0.0187 \pm 0.0013$.
Studying the antiferromagnet FeF$_2$, the authors of \cite{BeYo87} found 
$A_+/A_- = 0.53 \pm 0.01$ and $C_+/C_- = 4.6  \pm 0.02$.  
In particular for $A_+/A_-$ the accuracy of the experimental studies is 
close to ours. The results of both studies are consistent with ours, 
confirming universality.
For a comprehensive summary of experimental results see refs. 
\cite{ahp,PeVi02}.

\section{Summary and conclusions}
We have simulated the Blume-Capel model on the simple cubic lattice 
at $D=0.641$ and $0.655$
for a large number of inverse temperatures $\beta$ in a neighborhood
of the critical point. 
These values of $D$ are close to $D^*=0.656(20)$, where the amplitudes of 
leading corrections to scaling vanish.
We have simulated lattices up to $300^3$ in the 
high temperature and $500 \times 250^2$ in the low temperature phase.
Throughout we have chosen the linear size $L$ of the lattice such that 
$L \gtrapprox 10 \xi_{2nd}$ in the high temperature and 
$L \gtrapprox 20 \xi_{2nd}$ in the low temperature phase to avoid significant 
deviations from the thermodynamic limit. In the high temperature phase 
at $D=0.655$ we have reached the correlation length $\xi_{2nd} = 26.698(7)$.
Using the data obtained in these simulations we have extracted precise numerical
estimates for a number of universal amplitude ratios. We carefully estimated
systematical errors caused by subleading corrections. 

In table \ref{comparelattice} we have summarized our results and compare
them with previous estimates obtained from Monte Carlo simulations 
or from high and low temperature series expansions of lattice models. 
Our results are essentially consistent with but more precise than 
previous estimates. The same holds for the comparison with field theoretic
methods.  Also the accuracy of experimental results given in the 
literature is lower than ours.

\section{Acknowledgements}
This work was supported by the DFG under the grant No HA 3150/2-1.

\end{document}